\documentclass{article}

\PassOptionsToPackage{numbers, compress}{natbib}
\usepackage[preprint]{neurips_2026}

\usepackage[utf8]{inputenc}
\usepackage[T1]{fontenc}
\usepackage{amsmath}
\usepackage{amssymb}
\usepackage{amsfonts}
\usepackage{booktabs}
\usepackage{capt-of}
\usepackage{graphicx}
\usepackage{microtype}
\usepackage{nicefrac}
\usepackage{url}
\usepackage{wrapfig}
\usepackage{xcolor}
\usepackage{xspace}

\usepackage{hyperref}
\hypersetup{
    colorlinks=true,    
    urlcolor=cyan,
}
\usepackage[capitalize]{cleveref}

\usepackage{bm}
\usepackage{gensymb}
\usepackage{enumitem}

\newcommand{\eg}{\textit{e.g.}\xspace}

\newcommand{\minorsection}[1]{\noindent\textbf{#1}}

\definecolor{Highlight}{HTML}{39b54a}  

\usepackage{pifont}
\crefname{section}{Sec.}{Secs.}
\Crefname{section}{Section}{Sections}
\crefname{table}{Tab.}{Tabs.}
\Crefname{table}{Table}{Tables}
\crefname{figure}{Fig.}{Figs.}
\Crefname{figure}{Figure}{Figures}

\newcommand{\methodname}{\text{FFAvatar}\xspace}
\newcommand{\modelname}{\text{FFAvatar}\xspace}

\title{FFAvatar: Few-Shot, Feed-Forward, and Generalizable Avatar Reconstruction}

\author{
Thuan Hoang Nguyen$^{1,3}$\quad
Jiahao Luo$^{1,2}$\quad
Yinyu Nie$^{1}$\\
\textbf{Hao Li}$^3$\quad
\textbf{Gordon Guocheng Qian}$^{1\dagger}$ \quad
\textbf{Jian Wang}$^{1\dagger}$
\\
\normalsize $^1$ Snap Inc.\quad $^2$ University of California, Santa Cruz \quad $^3$ MBZUAI  \\
\\
\href{https://ffavatar.github.io}{\texttt{\textcolor{purple}{Project Page: \underline{https://ffavatar.github.io}}}}
}

\begin{document}

\maketitle
\footnotetext[2]{Corresponding authors. Jian Wang initiated this project.}

\begin{center}
    \centering
    \includegraphics[width=\linewidth]{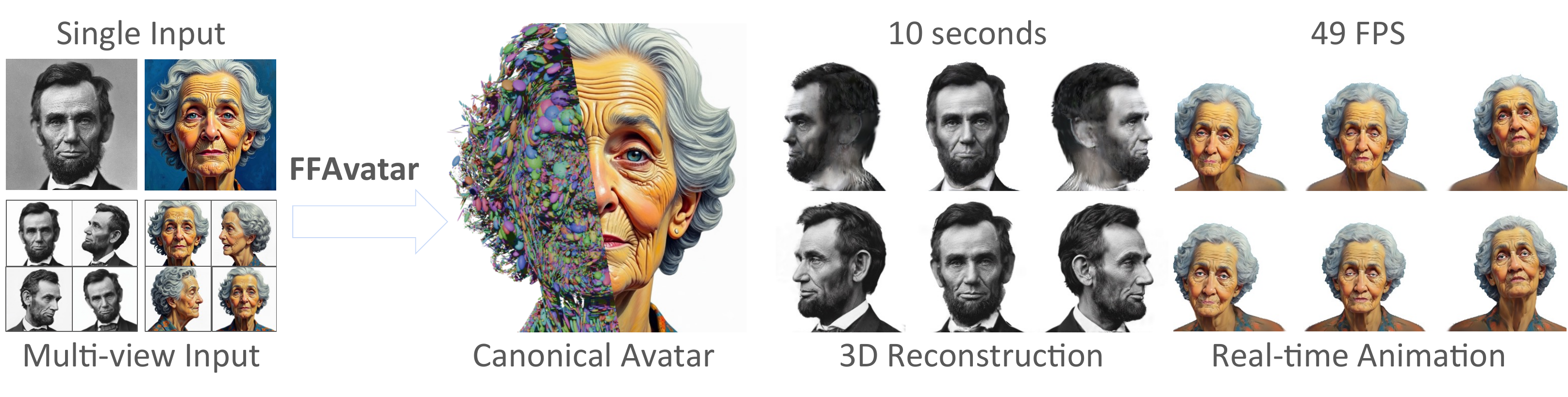}
    \captionof{figure}{
   FFAvatar full pipeline reconstructs animatable avatars in 10 seconds on a single A100, while supporting reenactment from driving frames at 49 FPS. Top: single-view; bottom: multi-view. 
    }
    \label{fig:teaser}
\end{center}

\begin{abstract}
Avatar reconstruction has traditionally relied on per-subject optimization that requires hours of computation or on expensive preprocessing that limits scalability. We introduce FFAvatar, a generalizable feed-forward framework that reconstructs high-quality, animatable 3D Gaussian head avatars from few-shot unposed portrait images in seconds. FFAvatar fuses information from multiple source images into a unified canonical Gaussian representation through  Multi-View Query-Former, which is animated via FLAME parameters predicted end-to-end directly from pixels, eliminating the overhead of offline FLAME extraction. We further propose a three-stage training curriculum that achieves both broad generalization and high-fidelity reconstruction: (i) scalable pretraining on extensive monocular video data with over 1M identities to learn strong generalizable priors; (ii) multi-view fine-tuning on a small but high-quality dataset of 360-degree captures to enhance geometric fidelity and extreme-view awareness; and (iii) optional personalization that adapts to specific identities for maximum fidelity within 500 optimization steps. Extensive experiments demonstrate that FFAvatar sets a new standard for identity preservation, geometric consistency, and animation fidelity. On the NeRSemble benchmark, it outperforms the state-of-the-art LAM by a substantial 5.5 PSNR gain. Furthermore, FFAvatar enables real-time deployment, reconstructing avatars in 2 seconds without personalization and 10 seconds with personalization, while supporting 49 FPS animation on a single NVIDIA A100 GPU.
\end{abstract}

\section{Introduction}\label{sec:intro}

\begin{figure*}[!t]
    \centering
    \includegraphics[width=\linewidth]{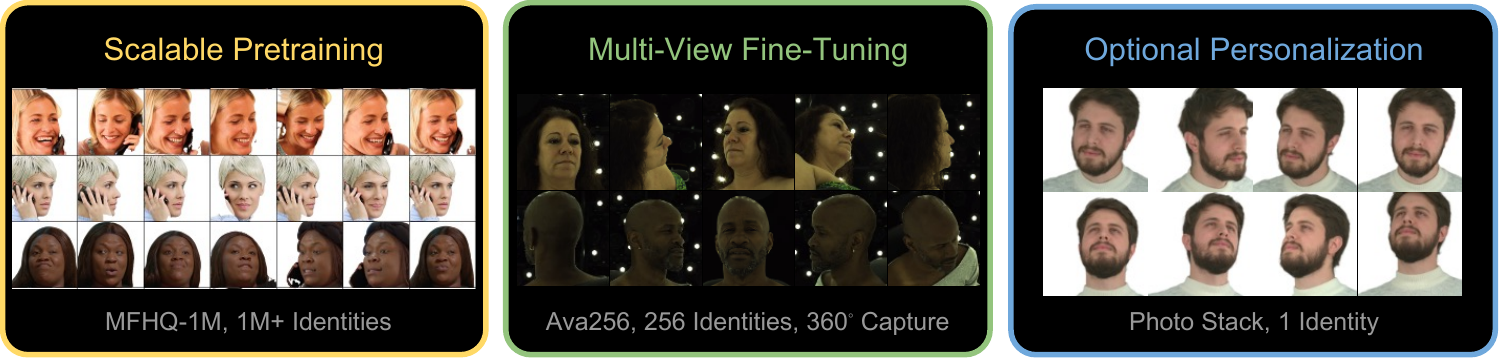}
    \caption{\textbf{Three-stage training} of \methodname{}. 
    Scalable pretraining fosters generalization across unseen identities by training on our private large-scale multi-frame-per-identity dataset MFHQ-1M, multi-view fine-tuning enhances geometric fidelity by optimizing the pretrained weights on a small-scale set of 360$\degree$ multi-view captures (\eg Ava256~\citep{martinez2024codec}), and lightweight personalization efficiently improves identity preservation with a few hundred tuning steps for a target identity in $<7$ seconds on a single A100 GPU.
    }
    \label{fig:training}
\end{figure*}

Recent progress in neural 3D avatar reconstruction \citep{gafni2021dynamic,zheng2022imavatar,zielonka2023instant} has produced high-quality digital humans, yet these methods remain bottlenecked by per-subject optimization that requires hours of computation and dozens to hundreds of images per identity. This fundamental limitation restricts their utility in practical applications where rapid deployment and minimal subject-specific data are paramount, such as virtual presence and telepresence. 

The recent Large Avatar Model (LAM) \citep{he2025lam} marks a significant advance by eliminating per-subject optimization: it predicts animatable 3D Gaussian avatars in a single feed-forward pass, achieving unprecedented inference speed across identities. However, \textit{LAM has two critical limitations}. First, it operates on single-view inputs, which constrains identity preservation and geometric fidelity, particularly for unseen or extreme viewpoints where regions are occluded or poorly observed in the input. This missing information must therefore be hallucinated by the model, leading to reduced fidelity. Second, LAM depends on expensive precomputed FLAME~\citep{li2017flame} parameter extraction, which fundamentally limits its scalability to training on large, unconstrained datasets and thus degrades the generalization of the final model.

We introduce \textbf{\methodname}, a framework that addresses both limitations by reconstructing animatable 3D head avatars from \textbf{multiple unposed portrait images} in a single feed-forward pass for any unseen identity (\cref{fig:teaser}) through a \textbf{multi-stage training} strategy (\cref{fig:training}).

Achieving this level of generalization is nontrivial due to a fundamental dataset dilemma. One could train directly on high-quality 360-degree capture datasets, but these are severely limited in diversity. One of the largest available datasets is Ava256~\citep{martinez2024codec}, which contains only 256 identities, causing models to overfit and fail to generalize to unseen identities at inference (see \cref{fig:ablation}). Conversely, large-scale in-the-wild video datasets offer abundant frames across identities but lack true multi-view coverage and 360-degree geometric supervision. This motivates our \textbf{first key contribution}: a \textbf{three-stage training curriculum}. As illustrated in \cref{fig:training}, we begin with \textbf{scalable pretraining} on diverse videos containing numerous identities, where multiple frames of the same person provide varied expressions and viewpoints. Although not truly 360-aware, this stage establishes strong generalization across identities. We then perform \textbf{multi-view fine-tuning} on small but high-quality multi-view datasets to inject geometric fidelity and 360-degree awareness; because the model is already pretrained, we find that even a modest dataset like Ava256~\citep{martinez2024codec} suffices to impart multi-view consistency. Finally, we support \textbf{optional personalization}, where our model can rapidly adapt to specific identities in fewer than 500 steps and 7 seconds on one A100 GPU, dramatically faster than optimization-based methods that must train from scratch.

Beyond data challenges, previous state-of-the-art methods \citep{he2025lam,xu2023gaussianheadavatar} rely on camera calibration or external FLAME parameter estimation, which requires expensive preprocessing pipelines. Applying such preprocessing at the scale needed for our pretraining stage would be prohibitively costly under computational budgets. This preprocessing bottleneck fundamentally limits scaling to large, unconstrained datasets. Our \textbf{second key contribution} addresses this limitation by learning a \textbf{FLAME Estimator end-to-end in a self-supervised manner}: we predict per-view expressions and poses directly from raw pixels through photometric supervision, eliminating external preprocessing and enabling scalable, robust avatar reconstruction as well as streaming avatar animation. 

Our \textbf{third key contribution} is the \textbf{multi-view architecture and few-to-many training objective} that enables \methodname{} to reconstruct a single, unified canonical Gaussian representation from multiple unposed input images. Unlike prior single-view methods, our architecture processes all input views jointly: image features from multiple viewpoints are aggregated into the 3D queries from FLAME canonical vertices, producing a consistent set of canonical Gaussian splats. By fusing information across multiple viewpoints, our approach achieves superior identity preservation and geometric consistency. 
\methodname{} is trained with a few-to-many objective: at each step, the model consumes a small conditioning subset of views to reconstruct the canonical avatar, then renders a larger set of target views with different expressions and poses. This training strategy teaches the model to generalize to unseen expressions and viewpoints of the same identity, ensuring robust performance even when only a few images are available at inference.

We summarize our \textbf{contributions} as follows:
\begin{itemize}[leftmargin=*,noitemsep,topsep=0pt,parsep=0pt,partopsep=0pt]
    \item \textbf{Three-stage training curriculum:} A progressive strategy for broad generalization and high-fidelity reconstruction via scalable pretraining, multi-view fine-tuning, and optional personalization.
    \item \textbf{End-to-end FLAME estimation:} A learnable FLAME Estimator trained end-to-end to predict FLAME parameters directly from pixels, eliminating external preprocessing for scalable training.
    \item \textbf{Multi-view avatar framework:} A generalizable feed-forward architecture with a few-to-many objective for reconstructing animatable 3D Gaussian head avatars from sparse unposed views. Extensive experiments demonstrate state-of-the-art performance of \methodname{} in generalization, geometric fidelity, and animation quality on various benchmarks.
\end{itemize}

\section{Related Work}
\label{sec:related_work}

\minorsection{Optimization-Based Avatar Reconstruction}
Traditional avatar reconstruction methods rely on per-subject optimization to fit parametric head models \citep{theis2019face2face} or neural representations \citep{gafni2021dynamic,hong2021headnerf} to multi-view captures or monocular videos. NeRF-based head avatar methods \citep{gafni2021dynamic,Magic123,hong2021headnerf,zheng2022imavatar} achieve high-quality, photorealistic results by optimizing implicit neural representations, often with explicit 3D priors or tracked FLAME parameters. However, these methods require hours to days of optimization per identity, along with dozens to hundreds of input frames or calibrated multi-view captures. Recent work has extended neural avatar reconstruction to 3D Gaussian Splatting representations \citep{qian2024gaussianavatars,xu2023gaussianheadavatar,zielonka2023instant,shao2024splattingavatar}, which enable real-time rendering and improved geometric detail. These methods remain strong optimization-based baselines, but their training-time data and computation requirements differ substantially from our few-shot feed-forward setting. While optimization-based approaches produce high-quality results, their computational demands limit scenarios where users may provide only a few images and cannot wait for lengthy per-subject processing.

\minorsection{Feed-Forward Avatar Reconstruction}
To overcome the computational bottleneck of optimization-based methods, recent work has explored feed-forward approaches that predict avatars in a single forward pass. Early encoder-decoder methods \citep{feng2021deca,danecek2022emoca,khakhulin2022rome} leverage parametric priors such as 3DMM~\citep{blanz2023morphable} or FLAME~\citep{li2017flame} to enable single-view reconstruction, but they lack photorealism or focus on 3D face understanding rather than synthesis. GPAvatar~\citep{chu2024gpavatar} reconstructs generalizable head avatars from one or several images using a dynamic point-based expression field and multi-triplane attention, but it predates recent Gaussian large-avatar models and is not the strongest public baseline for our NeRSemble setting. More recent approaches leverage large-scale transformer architectures and foundation models for improved generalization. GAGAvatar~\citep{chu2024gagavatar} introduces a dual-lifting mechanism that combines 2D image features with 3DMM-guided expression control, enabling animatable avatar generation from a single image. Avat3r~\citep{kirschstein2023avat3r} extends Large Reconstruction Models (LRMs)~\citep{hong2024lrm} to avatar reconstruction by incorporating DUSt3R~\citep{wang2024dust3r} dense correspondence and Sapiens~\citep{khirodkar2024sapiens} human-centric features to stabilize multi-view 3D lifting, but remains limited to expressions present in its training dataset and cannot generalize to arbitrary novel expressions. The recent Large Avatar Model (LAM)~\citep{he2025lam} represents a significant breakthrough by training on large-scale data to achieve unprecedented generalization across identities. LAM predicts canonical 3D Gaussian splats from a single image through a transformer architecture, enabling immediate reenactment via learned linear blend skinning weights. However, LAM has two critical limitations that restrict practical deployment: single-view input and precomputed FLAME parameters. Our work addresses both by extending large-scale avatar models to multi-view inputs and removing external FLAME preprocessing.

\section{Methodology}
\label{sec:method}

\begin{figure*}[!t]
\centering
\includegraphics[
  width=0.95\linewidth,
  trim=400 445 475 215,
  clip
]{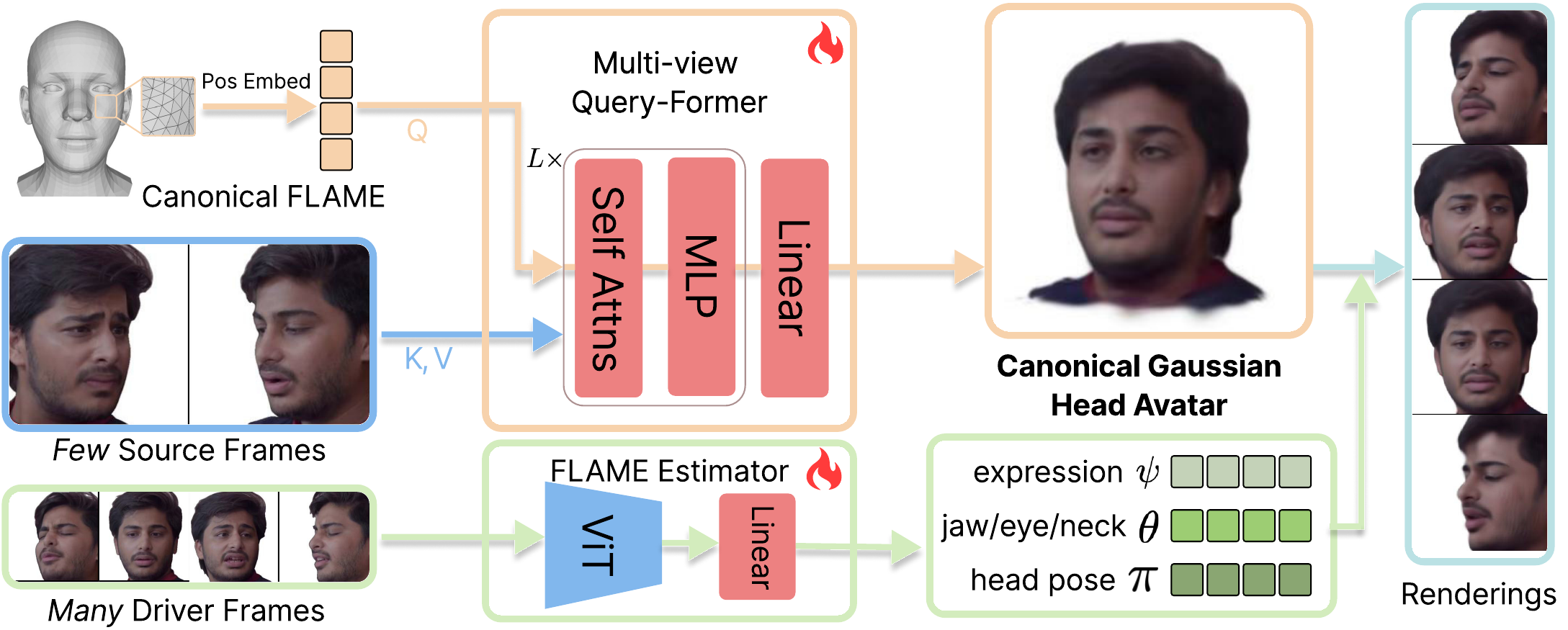}
\caption{\textbf{\methodname{} pipeline.}
\methodname{} reconstructs a canonical Gaussian head avatar from few-shot views using a Multi-view Query-Former, with canonical FLAME vertices as queries and source features as keys/values. An end-to-end FLAME Estimator predicts expression $\psi$, local articulation $\theta$, and head pose $\pi$ from driving frames, avoiding offline FLAME preprocessing. A few-to-many objective further improves generalization to unseen expressions and poses.
}
    \label{fig:pipeline}
\end{figure*}

\noindent We introduce \modelname{}, a multi-view large avatar model that reconstructs an animatable 3D head avatar directly from few-shot \textbf{unposed} portrait images. \modelname{} (i) proposes a \emph{multi-view} Query-Former that fuses information across multiple input images, and (ii) learns a \textbf{FLAME Estimator end-to-end} to remove the need for expensive FLAME preprocessing. \methodname{} avoids camera calibration and offline FLAME tracking, making it scalable for large-scale training.
We further introduce a three-stage training curriculum for optimizing this generalizable, animatable, and high-fidelity 3D avatar reconstruction model. 

\subsection{Preliminary}
\label{sec:preliminary}

\minorsection{Problem Formulation and Notation.}
Given \(N\) images \(\{I_n\}_{n=1}^{N}\) of a single identity captured under arbitrary viewpoints and expressions, our goal is to reconstruct a 3D head avatar represented as a set of \(M\) Gaussian splats in canonical space:
\begin{equation}
    \mathcal{G}^{\mathrm{can}} = \{\mu_m,\Sigma_m,\alpha_m,c_m\}_{m=1}^{M},
\end{equation}
\noindent with center \(\mu_m\in\mathbb{R}^3\), positive-definite covariance \(\Sigma_m\in\mathbb{R}^{3\times3}\), opacity \(\alpha_m\in(0,1)\), and color \(\mathbf{c}_m\). Here, we set \(\mu_m = v_m + o_m \), where \(v_m\) is a canonical vertex of the FLAME template and \(o_m\in\mathbb{R}^3\) is a learnable local offset predicted by the model for the target identity.
Throughout the paper, \(S\) denotes the number of conditioning source images, \(R\) denotes the number of reconstruction or driving images used for supervision, and \(T\) denotes the number of image tokens per view.


\paragraph{FLAME prior.} Li {\it et al.} ~\cite{li2017flame} represents a head using three \textit{fixed} and largely disentangled
sets of blendshape templates: identity, expression, and local articulation. The coefficients
$\beta$, $\psi$, and $\theta$ act as blending weights for the identity, expression, and articulation templates,
respectively. We use this structure to separate identity from animation: identity-specific geometry and appearance are modeled by the canonical Gaussian avatar $\mathcal{G}^{\mathrm{can}}$, whose
Gaussians are anchored to canonical FLAME vertices, while expression and pose are handled by
FLAME controls. 

\subsection{Multi‑View Large Avatar Model (\modelname{})}
\modelname{} is a fully \textbf{multi-view} framework that jointly aggregates information across multiple unposed portrait images, as shown in \cref{fig:pipeline}. 
Instead of processing each image independently, \modelname{} introduces a \textbf{Query-Former (Q-Former)}~\citep{li2023blip} module that performs geometry-aware cross-attention from canonical 3D queries to all image tokens from multiple views. 
This mechanism fuses complementary cues—such as geometry and textures from complementary views—into a consistent canonical representation. 
In addition, we train a \textbf{FLAME Estimator} end-to-end that predicts per-view FLAME parameters \((\psi,\theta,\pi)\) directly from image embeddings for animation, removing the need for any external FLAME preprocessing or camera calibration. Here, \(\psi\) denotes FLAME expression coefficients, \(\theta\) denotes local articulation parameters for jaw, eyes, and neck, and \(\pi=(R_h,t_h)\) denotes the global head pose applied to the avatar in a normalized camera frame.


\minorsection{FLAME Estimator \(\mathcal{F}\).}
Each driver image \(I_r\) is encoded by a ViT~\citep{ViT} (initialized from DINOv2~\citep{oquab2024dinov2}) into tokens \(F_r\in\mathbb{R}^{T\times C}\) and then put through a lightweight MLP head \(f_{\text{per‑view}}\) to infer per‑view FLAME parameters:
\begin{equation}
(\psi_r,\theta_r,\pi_r) = \mathcal{F}(I_r) = f_{\text{per‑view}}(F_r),
\end{equation}
This FLAME estimator $\mathcal{F}$ stays meaningful by predicting the identity-disentangled LBS weight $\psi,\theta,\pi$, restricted to only blending the fixed FLAME templates. As a result, the canonical Gaussian avatar \(\mathcal{G}^{\mathrm{can}}\) can also be driven by explicit FLAME parameters from any external tracker.

\minorsection{Multi-view Query-Former $\mathcal{D}$.} For the conditioning set \(\{I_s\}_{s=1}^{S}\), frozen DINOv2 extracts one feature sequence per view, \(F_s\in\mathbb{R}^{T\times C_{\mathrm{in}}}\). We concatenate tokens along the sequence dimension and apply a shared channel projection:
\begin{equation}
    F = W_F\!\left(\operatorname{concat}(F_1,\ldots,F_S)\right)\in\mathbb{R}^{ST\times C},
\end{equation}
where $W_F \in \mathbb{R}^{C_{in} \times C}$ projects input features concatenated from variable input views. 

For the input queries, we instantiate one projected learnable query per Gaussian/FLAME vertex, \(q_m=W_q(\phi(v_m))\), giving \(Q = \{q_m\}_{m=1}^M\in\mathbb{R}^{M\times C}\), where $\phi$ denotes positional embedding and $v_m$ denotes the canonical vertex of the FLAME template. The \(L\)-block Query-Former \(\mathcal{D}\) performs self-attention over the fixed-size query set and cross-attention from \(Q^{M\times C}\) to the variable-length multi-view token bank \(F^{ST\times C}\), outputting \(M\) updated tokens which are decoded as identity-injected avatar in canonical space $\mathcal{G}^{\mathrm{can}}$. This multi-view Query-Former process is formulated as:
\begin{align}
    \{o_m,\Sigma_m,\alpha_m,c_m\}_{m=1}^{M} = \mathcal{D}(\{v_m\}, F), \\
    \mathcal{G}^{\mathrm{can}} = \{v_m + o_m,\Sigma_m,\alpha_m,c_m\}_{m=1}^{M}
\end{align}

\minorsection{Animation.} Each Gaussian is anchored to one vertex of the canonical FLAME template. To enable animation, we deform only the Gaussian center and keep its covariance, opacity, and color unchanged. Given expression $\psi_r$, pose $\theta_r$, and global head pose $\pi_r$ for driving frame $r$, FLAME linear blend skinning provides the blended transform
\[
(R_{m,r},t_{m,r})=\sum_b w_{m,b}A_{b,r}(\psi_r,\theta_r,\pi_r),
\]
where $w_{m,b}$ is the fixed FLAME skinning weight of the anchor vertex and \(A_{b,r}\) is the FLAME bone transform. The Gaussian center is then deformed as $\mu'_{m,r}=R_{m,r}\mu_m+t_{m,r}.$


\subsection{Training Objectives}
\label{sec:losses}

$\mathcal{F}$ and $\mathcal{D}$ are optimized end-to-end through differentiable rendering losses after FLAME-based animation by the  few-to-many objective as follows. 

\minorsection{Few‑to‑Many Objective.} At each training iteration, given the complete image set of an identity, we randomly select two \emph{disjoint} subsets: a conditioning subset \(\{I_s\}_{s=1}^{S}\) and a reconstruction subset \(\{I_r\}_{r=1}^{R}\) with \(R \ge S\). While previous works focus on reconstructing a single target view from one or multiple inputs~\citep{chu2024gagavatar,he2025lam}, our few-input, many-target objective aligns with the goal of avatar reconstruction: using a small number of input views to learn an avatar that can be rendered from arbitrary viewpoints.
The canonical avatar decoder \(\mathcal{D}\) consumes only the conditioning views to predict the canonical Gaussian splats, while the FLAME Estimator \(\mathcal{F}\) predicts the FLAME parameters for each reconstruction view:
\begin{align}
\mathcal{G}^{\mathrm{can}} = \mathcal{D}\big(\{I_s\}_{s=1}^{S}\big), \quad
\{\psi_r,\theta_r,\pi_r\}_{r=1}^{R}=\mathcal{F}\big(\{I_r\}\big).
\end{align}
For each target view \(r\), we deform \(\mathcal{G}^{\mathrm{can}}\) via linear blend skinning (LBS)~\citep{li2017flame} and render the output under the normalized camera:
\[
\mathcal{G}^{\mathrm{pose}}_r=\mathrm{LBS}\!\big(\mathcal{G}^{\mathrm{can}},\psi_r,\theta_r,\pi_r\big), 
\qquad 
\widehat{I}_r=\mathcal{R}\!\big(\mathcal{G}^{\mathrm{pose}}_r\big).
\]
Losses are computed over all \(\{I_r\}_{r=1}^{R}\).
By this scheme, the model learns to generalize from \textbf{few} conditioning views \(\{I_s\}_{s=1}^{S}\) to \textbf{many} reconstruction targets \(\{I_r\}_{r=1}^{R}\).

\minorsection{Photometric losses.}
The rendered RGB images are supervised using a combination of photometric and perceptual losses computed with respect to their corresponding ground‑truth images:
\begin{align}
\mathcal{L}_{1} &= \sum_r \|I_r - \widehat{I}_r\|_{1}, \quad 
\mathcal{L}_{\mathrm{lpips}} = \sum_r \mathrm{LPIPS}(I_r, \widehat{I}_r), \quad 
\mathcal{L}_{\mathrm{ssim}} = \sum_r \big(1 - \mathrm{SSIM}(I_r, \widehat{I}_r)\big).
\end{align}

\minorsection{Adversarial loss.}
Training with only pixel and perceptual supervision often produces overly smooth results. 
To enhance texture fidelity and realism, we introduce an adversarial loss 
$\mathcal{L}_{\mathrm{adv}}$ employing a projected discriminator~\citep{sauer2021projectedgan} 
with differentiable augmentation~\citep{zhao2020diffaugment}. 
Unlike prior feed-forward avatar reconstruction approaches such as 
GAGAvatar~\citep{chu2024gagavatar} and LAM~\citep{he2025lam}, 
we incorporate adversarial supervision into our framework, 
which improves texture sharpness and overall rendering quality.

\minorsection{Total loss.} The total training loss is a weighted combination of all the terms mentioned above:
\begin{equation}
\mathcal{L} = 
\lambda_{1}\mathcal{L}_{1} +
\lambda_{2}\mathcal{L}_{\mathrm{lpips}} +
\lambda_{3}\mathcal{L}_{\mathrm{ssim}} +
\lambda_{4}\mathcal{L}_{\mathrm{adv}},
\end{equation}
where we set \(\lambda_1 = 0.8\), \(\lambda_2 = 0.1\), \(\lambda_3 = 0.1\), and \(\lambda_4 = 0.01\) empirically.

\begin{figure*}[t]
    \centering
    \includegraphics[width=\linewidth]{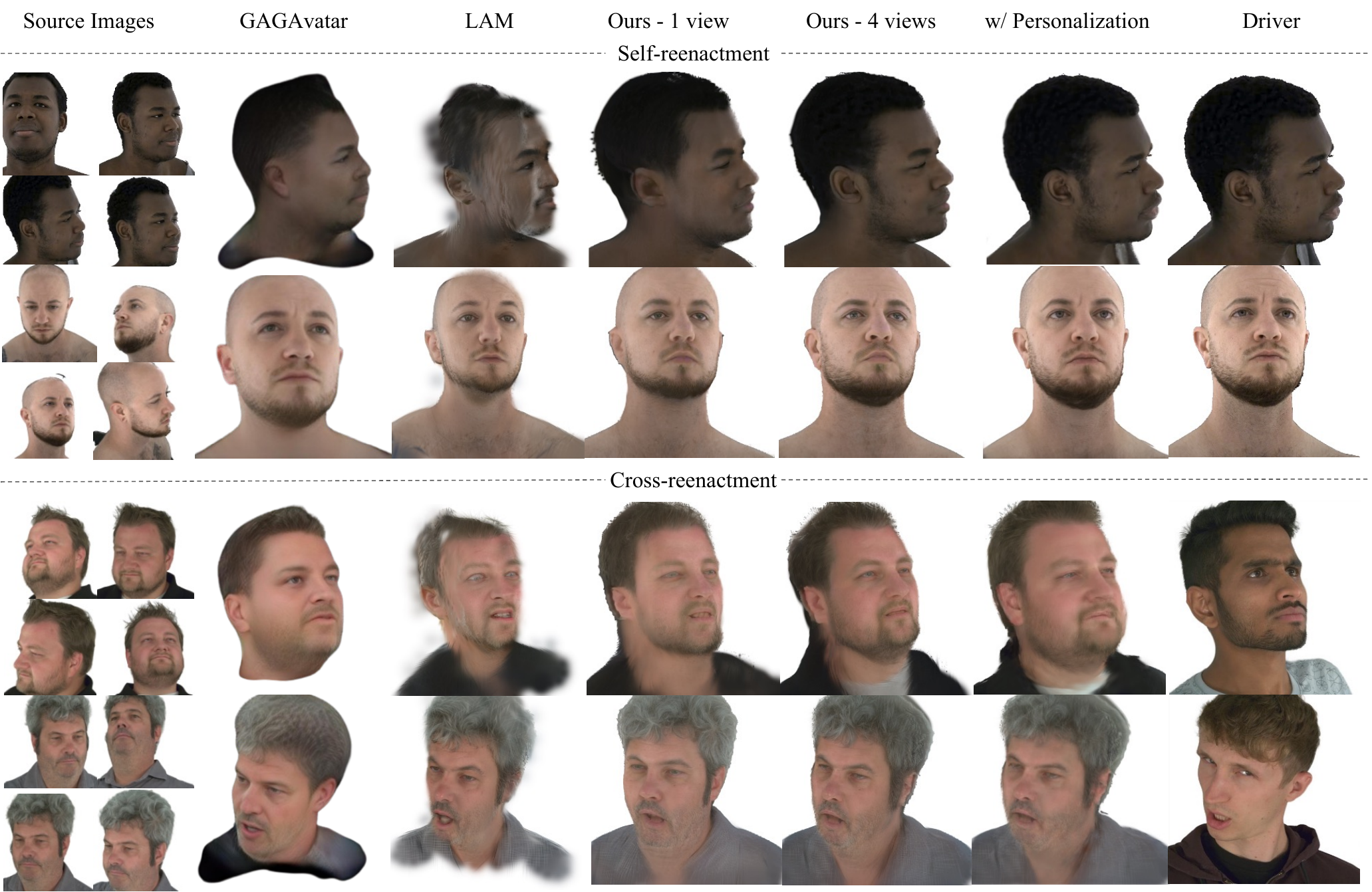}
    \captionof{figure}{\textbf{\methodname{} qualitative comparison} for self-reenactment on the \textbf{Ava256} test set (top two rows) and cross-reenactment on the \textbf{NeRSemble} benchmark (bottom two rows).
    \methodname{}-1 view achieves more faithful and geometrically consistent results than the baselines. GAGAvatar~\citep{chu2024gagavatar} often produces over-smoothed textures and pose misalignment, while LAM~\citep{he2025lam} shows geometry artifacts under challenging views. Additional input views and optional personalization further improve identity preservation and detail.
    }
    \label{fig:results}
\end{figure*}

\subsection{Training Strategy}
As illustrated in \cref{fig:training}, we propose a three-stage training strategy designed to progressively enhance generalization, geometric fidelity, and identity preservation through scalable pretraining, multi-view fine-tuning, and optional personalization. 

\minorsection{Scalable Pretraining.} We pretrain \modelname{} on large collections of easily accessible monocular videos, where multiple frames of the same identity naturally provide diverse expressions and viewpoints, as shown in \cref{fig:training} left. Consequently, this stage involves significantly more identities and longer training time than the subsequent stages. The goal is to build a strong prior that generalizes across identities. However, since most video sequences are monocular and not truly multi-view aware, we introduce a second stage that fine-tunes the model on high-quality multi-view captures to improve geometric fidelity and view consistency.

\minorsection{Multi-View Fine-Tuning.}  
High-quality 3D avatar reconstruction ultimately requires at least \(180\degree\) coverage to model 3D geometry. Collecting such data demands professional multi-view capture setups, making these datasets relatively scarce. We therefore reserve this data for a second-stage refinement phase (\cref{fig:training} middle), designed to further enhance cross-view consistency and geometric fidelity of the pretrained model from the scalable pretraining stage. During training, views are randomly sampled across all available camera angles to encourage full \(360\degree\) coverage and robustness to diverse viewpoints.

\minorsection{Optional Personalization.} For target subjects (multi-view collections of a single identity shown in \cref{fig:training} right), we propose an optional lightweight personalization stage. Learnable \textit{residuals} $\Delta_{\mathcal{G}^{\mathrm{can}}}$ on Gaussian attributes are optimized per subject with the Gaussians from the feed-forward model as initialization. The Gaussian parameters after personalization are formulated as: 
\begin{equation}
    \mathcal{G}^{\mathrm{can}}_{p} = \mathcal{G}^{\mathrm{can}} + \Delta_{\mathcal{G}^{\mathrm{can}}}.
\end{equation}
This stage efficiently enhances identity-specific details and typically converges in 500 optimization steps, which is $60\times$ faster than training from scratch that usually requires around 100K steps (\cref{fig:supp_progress}).

\section{Experiments}

\subsection{Experiment Setup}


\minorsection{Implementation Details.}
We first pretrain \methodname{} on our large-scale dataset MFHQ-1M for 200K steps. MFHQ-1M comprises 1M identities, each with 8 frames capturing diverse expressions and viewpoints sampled from monocular videos. For legal reasons, this dataset cannot be released. A similar dataset can be collected following Omni-ID~\citep{qian2025omni} and ComposeMe~\citep{qian2025composeme}. 
In the second stage, we fine-tune the pretrained weights on multi-view video captures from the \textbf{Ava256}~\citep{martinez2024codec} dataset for 20K steps. Specifically, we use the 4~TB version containing 7.5~fps recordings from 80 synchronized cameras (approximately 5{,}000 frames per subject). We use 248 identities for training and hold out the remaining 8 identities for evaluation. The third stage optimizes Gaussian residuals per identity for 500 steps. 
\(1\!-\!4\) images are randomly sampled as input in the first stage, and \(1\!-\!8\) images are used in the last two stages. For the reconstruction set size, we use 8, 16, and all available views in the three stages, respectively.

\methodname{} uses \(L=12\) blocks in the Multi-View Query-Former. The complete model contains 870.8M parameters, comprising 313.2M parameters in the FLAME estimator and 557.6M parameters in the avatar component. The whole pipeline is optimized using Adam~\citep{adam} with learning rates of \(10^{-5}\), \(10^{-6}\), and \(10^{-4}\) for the three stages and a batch size of 1. The first two stages are trained with 8~NVIDIA~A100~GPUs for 3 and 1.5 days, respectively, while the last stage uses one A100 GPU and takes only 7 seconds. The input and target resolutions are set to \(504\times504\) for all stages. 

Regarding the Gaussian avatar model, the original 5{,}023 FLAME vertices are insufficient for high-fidelity 3D Gaussian avatar reconstruction and are thus upsampled to 80K Gaussians following LAM~\citep{he2025lam}. For training efficiency, gradient checkpointing with \texttt{bfloat16} mixed precision is used.

\minorsection{Benchmark \& Metrics.}
We compare \methodname{} with state-of-the-art feed-forward head avatar methods, including GAGAvatar~\citep{chu2024gagavatar} and LAM~\citep{he2025lam}, using their official single-view reconstruction settings; Avat3r~\citep{kirschstein2023avat3r} is excluded because its code and checkpoint are unavailable. We evaluate generalization on the \textit{unseen} NeRSemble dataset, using 45 randomly selected identities with 16 camera views each, and report PSNR, SSIM, LPIPS~\citep{zhang2018unreasonable}, and ArcFace Cosine Similarity (CSIM)~\citep{deng2019arcface}. NeRSemble is challenging because its many side-view renderings require accurate 3D reconstruction, where prior methods such as LAM still leave substantial room for improvement. Since \methodname{} also supports feed-forward multi-view inputs, we further report multi-view results to highlight its scalability beyond the single-view setting.

\subsection{Results}
\minorsection{Qualitative Comparison.} A qualitative comparison between \methodname{} and baseline approaches is \begin{wraptable}[16]{r}{0.54\textwidth}
\centering
\renewcommand\tabcolsep{3pt}
\captionof{table}{\textbf{Quantitative comparison for self-reenactment on the NeRSemble benchmark.}
\methodname{} outperforms state-of-the-art feed-forward avatar methods~\citep{chu2024gagavatar,he2025lam} in the single-view setting. Using 4 input views further improves reconstruction by providing richer appearance and geometry cues, while 500-step personalization yields the best results through rapid identity-specific adaptation.
}
\label{tab:self_quant}
\resizebox{\linewidth}{!}{
\begin{tabular}{l|c|c|c|c}
    \toprule
    \textbf{Method} & 
    \textbf{PSNR}$\uparrow$ & 
    \textbf{SSIM}$\uparrow$ & 
    \textbf{LPIPS}$\downarrow$ & 
    \textbf{CSIM}$\uparrow$ \\
    \midrule
    GAGAvatar (1-view)~\citep{chu2024gagavatar} & 15.03 & 0.63 & 0.34 & 0.47 \\
    LAM (1-view)~\citep{he2025lam}      & 13.31 & 0.69 & 0.30 & 0.61 \\
    \textbf{Ours (1-view)}      & {17.60} & {0.75} & {0.23} & {0.68} \\
    \textbf{Ours (4-view)}      & {18.78} & {0.78} & {0.19} & {0.74} \\
    \textbf{Ours (4-view) + Personalization} & \textbf{25.78} & \textbf{0.94} & \textbf{0.09} & \textbf{0.89} \\
    \bottomrule
\end{tabular}
}
\end{wraptable}
shown in Fig.~\ref{fig:results}.  We evaluate both \textit{self-reenactment} (driver image from the same identity) and \textit{cross-reenactment} (driver image from a different identity). Table~\ref{tab:self_quant} reports the single-view setting used by GAGAvatar~\citep{chu2024gagavatar} and LAM~\citep{he2025lam}, our four-input setting, and the same four-input setting after optional personalization.
Qualitatively, GAGAvatar tends to produce overly smoothed geometry and exhibits noticeable head-pose misalignment, while LAM struggles with non-frontal or extreme poses, often generating holes and severe artifacts.
In contrast, \methodname{} mitigates head-pose ambiguity through the \textit{FLAME Estimator}, trained on large-scale video data during the scalable pretraining stage. Moreover, our multi-view fine-tuning on high-quality stereo captures enables sharper textures and more consistent geometry, effectively reducing over-smoothing and extreme-viewpoint artifacts, compared to LAM.
Notably, the 4-view input that only our \methodname{} supports achieves a significant improvement over the single-view setting, underscoring the importance of multi-view inputs for high-fidelity reconstruction. The optional personalization further enhances identity preservation as indicated in the column ``w/ Personalization''.

\begin{figure*}[!t]
    \centering
    \includegraphics[width=\linewidth]{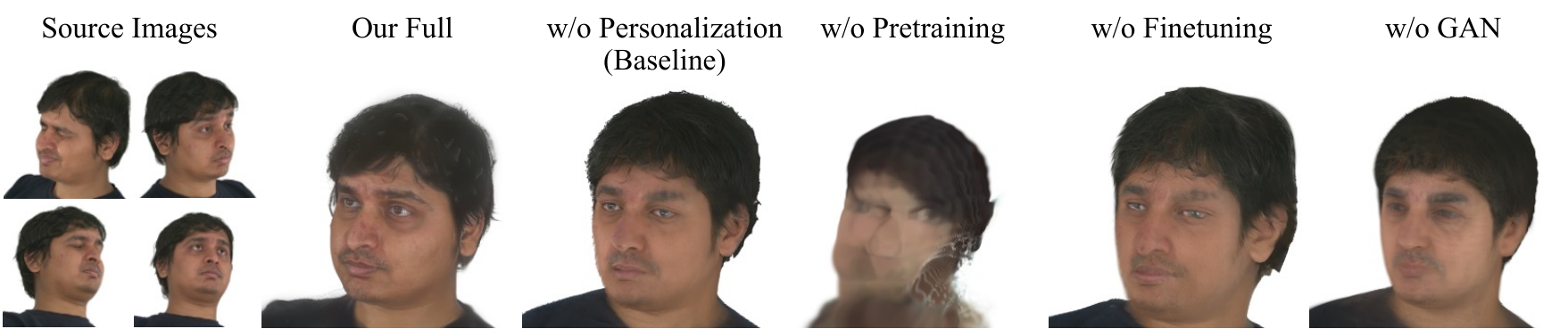}
\caption{\textbf{Qualitative ablation study.} \methodname{} with personalization achieves the most realistic and faithful reconstructions.
Personalization enhances identity. 
Without scalable pretraining, the model trained only on Ava256 fails to generalize to NeRSemble, degrading geometry and identity consistency. Removing high-quality fine-tuning or the GAN loss reduces \mbox{visual detail.}}
    \label{fig:ablation}
\end{figure*}

\minorsection{Quantitative Comparison.}
Following common practice, we report quantitative results on self-reenactment rendering in the \textbf{NeRSemble} benchmark.
As shown in Table~\ref{tab:self_quant}, \methodname{} achieves substantial improvements over existing feed-forward avatar reconstruction methods across all metrics, even in the single-view setting, e.g. +2.57 PSNR over the next-best method (GAGAvatar) and a notable gain in identity preservation (+0.21 CSIM). Compared to LAM, \methodname{} also delivers significantly higher rendering quality, as reflected in both PSNR and SSIM. Leveraging four input views further boosts performance (over +1 PSNR) and raises CSIM beyond 0.7, confirming that incorporating multi-view information significantly enhances geometry and appearance fidelity.
With optional personalization, reconstruction quality further improves across all metrics, demonstrating rapid adaptation to specific identities in only 500 steps and $7$ seconds on one A100 GPU.

\begin{wraptable}[8]{r}{0.54\textwidth}
\centering
\vspace{-2em}
\caption{\textbf{Quantitative ablation study.} 
}
\label{tab:ablation}
\renewcommand\tabcolsep{3pt}
\resizebox{\linewidth}{!}{
\begin{tabular}{l|c|c|c|c}
    \toprule
    \textbf{Method} & 
    \textbf{PSNR}$\uparrow$ & 
    \textbf{SSIM}$\uparrow$ & 
    \textbf{LPIPS}$\downarrow$ & 
    \textbf{CSIM}$\uparrow$ \\
    \midrule
    Ours (4-view)                 & \textbf{18.78} & \textbf{0.78} & \textbf{0.19} & \textbf{0.74} \\
    w/o GAN loss & 17.51 & 0.77 & 0.25 & 0.69 \\
    w/o scalable pretraining    & 10.42 & 0.64 & 0.46 & 0.39 \\
    w/o multi-view fine-tuning &  15.25 & 0.67 & 0.28 & 0.61\\
    w/o few-to-many loss & 15.91 & 0.74 & 0.24 & 0.71 \\
    \bottomrule
\end{tabular}
}
\end{wraptable}

\subsection{Ablation Study}
On the same NeRSemble test set, we ablate \textit{scalable pretraining}, \textit{multi-view fine-tuning}, the \textit{GAN} loss, and the \textit{few-to-many} loss.
As shown in Table~\ref{tab:ablation}, removing {scalable pretraining} leads to the largest drop (-8.36 PSNR, -0.35 CSIM), demonstrating that large-scale pretraining is crucial for robust generalization to unseen identities. Excluding {multi-view fine-tuning} also causes noticeable degradation (-3.53 PSNR, -0.13 CSIM), highlighting its role in refining geometry and texture details. Removing either the {GAN} loss or the few-to-many loss decreases perceptual realism, reflected in worse LPIPS and lower CSIM scores. 
Overall, these results validate the effectiveness of our three-stage training strategy in achieving both geometric consistency and visual fidelity. The qualitative comparisons in \cref{fig:ablation} further support these findings. 

\begin{table*}[t]
\centering
\scriptsize
\setlength{\tabcolsep}{3pt}
\caption{\textbf{Ablation: end-to-end FLAME estimator vs. state-of-the-art.}
Replacing our learned estimator with VHAP~\citep{qian2024vhap} keeps the avatar and personalization unchanged. Our estimator matches rendering quality while avoiding offline tracking and running over \(200\times\) faster.}
\label{tab:additional_analysis}
\vspace{0.35em}
\begin{minipage}{0.49\linewidth}
\centering
\textbf{(a) FLAME source with personalization}\\[-0.25em]
\resizebox{\linewidth}{!}{
\begin{tabular}{lccccc}
\toprule
\textbf{Method} & \textbf{PSNR}$\uparrow$ & \textbf{SSIM}$\uparrow$ & \textbf{LPIPS}$\downarrow$ & \textbf{CSIM}$\uparrow$ & \textbf{FPS}$\uparrow$ \\
\midrule
Ours & 25.78 & 0.94 & 0.09 & 0.89 & 60 \\
VHAP tracker & 25.81 & 0.96 & 0.09 & 0.90 & 0.3 \\
\bottomrule
\end{tabular}
}
\end{minipage}
\hfill
\begin{minipage}{0.49\linewidth}
\centering
\textbf{(b) FLAME coefficient difference vs. VHAP}\\[-0.25em]
\resizebox{\linewidth}{!}{
\begin{tabular}{lccccc}
\toprule
\textbf{Param.} & \textbf{Mean VHAP} & \textbf{Std VHAP} & \textbf{Mean Ours} & \textbf{Std Ours} & \textbf{RMSE} \\
\midrule
Expression & 0.034 & 0.011 & 0.027 & 0.014 & 0.012 \\
Jaw/eyes/neck & 0.012 & 0.002 & 0.036 & 0.004 & 0.020 \\
Head pose (rad.) & 0.34 & 0.43 & 0.48 & 0.39 & 0.17 \\
\bottomrule
\end{tabular}
}
\end{minipage}
\end{table*}

\minorsection{FLAME Estimator Analysis.}
\cref{tab:additional_analysis} validates the learned FLAME Estimator. Replacing our estimator with the state-of-the-art standalone VHAP tracker~\citep{qian2024vhap} gives on-par personalized rendering, showing that our LBS interface supports explicit FLAME coefficient driving from external trackers. Across 360K NeRSemble frames (500 frames, 16 views, 45 identities), coefficient differences from VHAP remain small. Our estimator runs at 60 FPS versus 0.3 FPS for offline VHAP tracking, over \(200\times\) faster without offline preprocessing.

\begin{wrapfigure}[16]{r}{0.58\textwidth}
    \vspace{-1em}
    \centering
    \includegraphics[width=\linewidth]{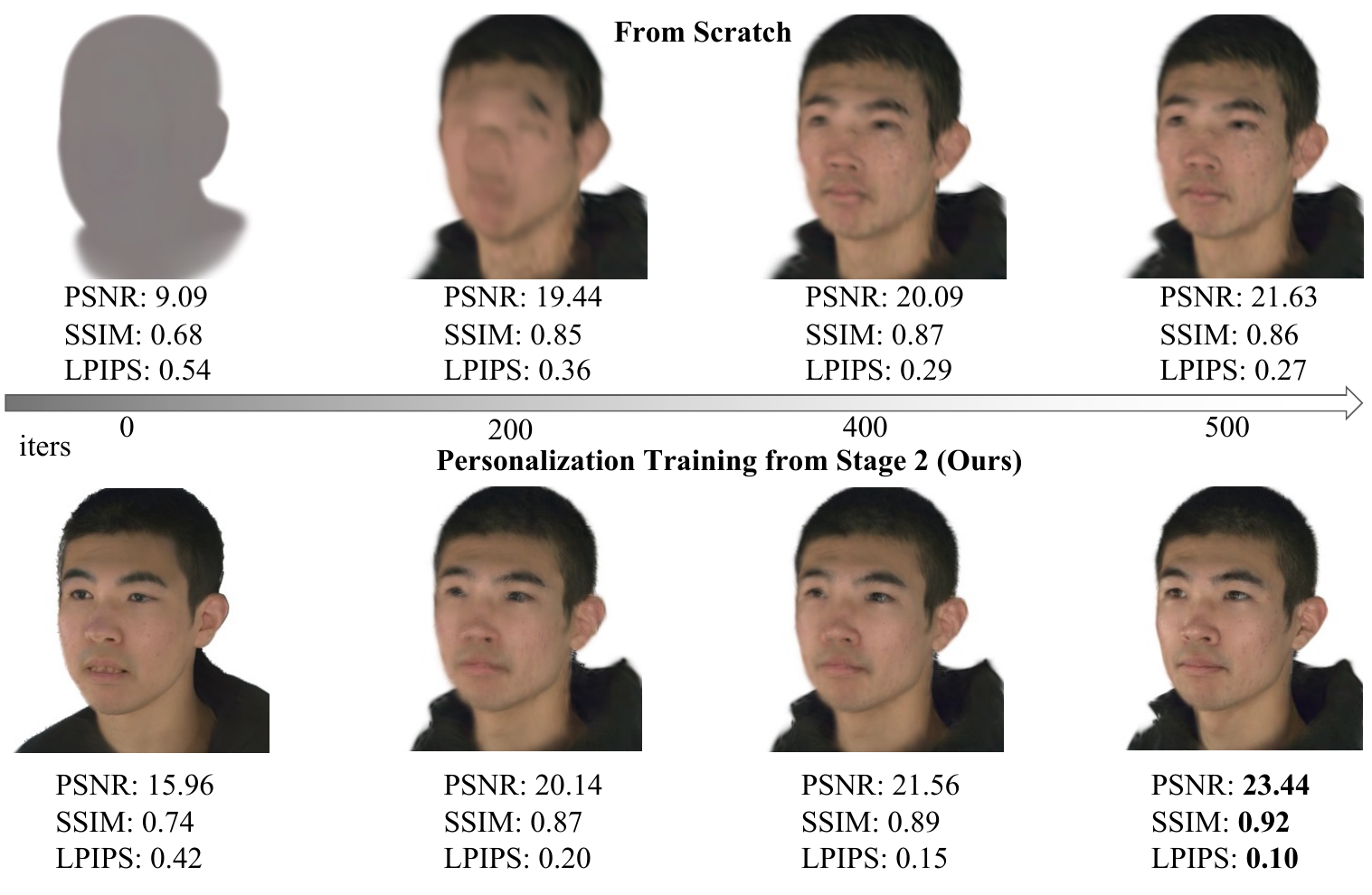}
    \vspace{-2em}
    \caption{\textbf{Personalization dynamics.}
    Feed-forward initialization improves quality and converges within 500 steps, while random initialization (from scratch), remains blurry and poorly preserves identity.}
    \label{fig:supp_progress}
\end{wrapfigure}

\minorsection{Personalization Analysis.}
\noindent\looseness=-1 We use 500 personalization steps because most examples converge by this point (\cref{fig:supp_progress}); longer optimization gives diminishing returns. Fig.~\ref{fig:supp_progress} also shows that feed-forward initialization provides a much better starting point than random initialization. Without personalization, sparse-view prediction averages high-frequency identity cues, so identity preservation is weaker; Gaussian-residual optimization restores subject-specific details efficiently.

\section{Limitation Analysis}
\label{sec:limitations}

While \methodname{} improves feed-forward avatar reconstruction, several limitations remain. Its animation prior is bounded by FLAME's solution space and thus lacks detailed modeling of eye gaze, mouth interior, and tongue geometry. Sparse input views may also miss hair, neck, or clothing boundaries, requiring hallucination that can introduce artifacts under extreme novel views. Finally, without personalization, single-step sparse-view prediction may smooth fine-grained identity details; optional Gaussian-residual personalization mitigates this with a small test-time cost.

\section{Conclusion}
\label{sec:conclusion}

We presented \methodname{}, a generalizable, feed-forward framework for reconstructing animatable 3D Gaussian head avatars directly from few-shot portrait images. By unifying scalable pretraining, multi-view fine-tuning, and optional lightweight personalization, \methodname{} achieves strong identity generalization, high subject-specific fidelity and geometric consistency across extreme viewpoints. During inference on a single NVIDIA A100 GPU, \methodname{} reconstructs an avatar in under 2 seconds with our feed-forward pipeline, including preprocessing. For enhanced fidelity, optional personalization can be completed in an additional 7 seconds. \methodname{} also achieves 49 FPS animation without precomputed FLAME parameters. We believe \methodname{} offers a scalable foundation for future research on controllable, real-time human avatar synthesis and serves as a step toward more accessible digital human creation.


{
    \small
    \bibliographystyle{plainnat}
    \bibliography{main}
}

\newpage

\appendix

\section{Experiment Setup Details}
\label{sec:supp_exp}

\minorsection{Baselines.}
We compare \methodname{} with state-of-the-art feed-forward head avatar generation methods including GAGAvatar~\citep{chu2024gagavatar} and LAM~\citep{he2025lam}. Avat3r~\citep{kirschstein2023avat3r} is not compared because its checkpoint and code are not available. All methods are evaluated under their official single-view reconstruction settings for fair comparison.
GPAvatar~\citep{chu2024gpavatar} and NeRF-based multi-view avatar methods~\citep{gafni2021dynamic,hong2021headnerf,zheng2022imavatar} are discussed in the main related work section but not included in the main quantitative table because their evaluation settings differ substantially: GPAvatar predates stronger recent feed-forward Gaussian baselines, while NeRF-based methods require per-subject optimization with many frames or calibrated captures. In contrast, \methodname{} additionally supports multi-view inputs in a feed-forward setting. To further highlight the benefits of this design, we also report \methodname{}'s performance under multi-view configurations and compare it against the single-view results.

\minorsection{Benchmark \& Metrics.}
To evaluate the generalization ability of our model, we conduct experiments on the \textit{unseen} \textbf{NeRSemble} dataset, testing its performance in reconstructing high-fidelity 3D head avatars for novel subjects from both single-image and multi-view inputs. We evaluate all methods on a NeRSemble test set consisting of 45 randomly selected identities, each captured under 16 camera views.
We adopt three standard paired-image metrics to assess rendering quality: Peak Signal-to-Noise Ratio (PSNR), Structural Similarity Index Measure (SSIM), and Learned Perceptual Image Patch Similarity (LPIPS)~\citep{zhang2018unreasonable}.
To further evaluate identity preservation, we report cosine face similarity score (CSIM), defined as the cosine similarity between ArcFace~\citep{deng2019arcface} embeddings of the ground-truth and predicted renderings.
Note that NeRSemble is a particularly challenging benchmark, as it includes many side-view renderings that demand high-fidelity 3D reconstruction, which is essential for 3D avatar evaluation. For example, although LAM~\citep{he2025lam} demonstrates strong performance in front-view renderings, there remains substantial room for improvement in side-view quality.



\end{document}